\documentclass{jpp}
\usepackage{color,graphicx}
\usepackage{mathrsfs}
\usepackage{hyperref}

\title[]{Nonlinear reversed shear Alfv\'en eigenmode saturation due to spontaneous zonal current generation}

\author{Shizhao Wei\aff{1}, Tao Wang\aff{1,2}, Ningfei Chen\aff{1} and Zhiyong Qiu\aff{1,2}\corresp{\email{zqiu@zju.edu.cn}}}

\affiliation{\aff{1}Institute for Fusion Theory and Simulation and Department of Physics, Zhejiang University, Hangzhou 310027, People's Republic of China
\aff{2}Center for Nonlinear Plasma Science and ENEA C. R. Frascati, Frascati, Italy}

\begin{document}

\maketitle

\begin{abstract}
General nonlinear equations describing reversed shear Alfv\'en eigenmode (RSAE) self-modulation via zero frequency zonal structure (ZFZS) generation
are derived using nonlinear gyrokinetic theory, which are then applied to study the spontaneous ZFZS excitation as well as RSAE nonlinear saturation.
It is found that both electrostatic zonal flow (ZF) and electromagnetic zonal current (ZC)
can be preferentially excited by finite amplitude RSAE, depending on specific plasma parameters.
The modification to local shear Alfv\'en wave continuum is evaluated using the derived saturation level of ZC, which is shown to play a comparable role in saturating RSAE with the ZFZS scattering.

\end{abstract}


\section{Introduction}
\label{sec:introduction}

The future tokamak-based fusion power plants and experimental devices towards this goal (e.g., the ITER \citep{KTomabechiNF1991}), are expected to operate
at steady state with a substantial non-inductive current fraction \citep{CGormezanoNF2007}, which generally renders a reversed shear scenario \citep{JHuangNF2020}.
In this circumstance, the reversed shear Alfv\'en eigenmode (RSAE, also known as Alfv\'en cascade) is frequently observed \citep{SSharapovPLA2001}
as driven unstable by the energetic particles (EPs) \citep{FZoncaPoP2002,WChenNF2014,AFasoliNF2007,LChenRMP2016}
in present day tokamaks, and is expected to play significant roles in future reactors \citep{TWangPoP2018},
in transporting fusion alpha particles to tokamak edge \citep{TWangPoP2019}, which has a deleterious effects on plasma self-heating, and may damage the plasma facing components \citep{RDingNF2015}.
RSAE is a branch of shear Alfv\'en wave (SAW) eigenmode localized radially
near the minimum of the safety factor $q$-profile \citep{SSharapovPoP2002}, which is denoted as $q_{\rm min}$.
The lowest order RSAE frequency in the incompressible limit, $\omega\simeq k_{\parallel}v_{\rm A}\simeq|n-m/q_{\rm min}|v_{\rm A}/R$,
reflects the sensitive dependence on  the instantaneous $q_{\rm min}$ value for  given toroidal/poloidal mode numbers $n$/$m$, and this feature can be used in $q$-profile measurement, i.e., MHD spectroscopy \citep{WChenNF2014,AFasoliPPCF2002}.
Here, $k_{\parallel}$ is the wave number parallel to the equilibrium magnetic field $\bf B$, $v_{\rm A}$ is the Alfv\'en speed, and $R$ is the major radius.
RSAEs are typically dominated by one or two poloidal harmonic, with the radial width $\propto\sqrt{q/(r^{2}_{0}q'')}$ with $r_{0}$ being the radial location of $q_{\rm min}$, and $q''\equiv\partial^2_{r}q$.
Despite the fairly good understanding of these linear physics, the nonlinear dynamics of RSAE still attracts recent research interest,
especially in view of the EP as well as thermal particle transport induced by the associated electromagnetic field perturbations \citep{TWangPoP2019,TWangNF2020,PShiNF2020,SWeiCPL2021}.
The transport rate is closely related with the perturbation amplitude \citep{LChenJGR1999,FZoncaJPCS2021};
and in reactor relevant cases with many SAWs simultaneously driven unstable by EPs, the EP orbit could become chaotic
and eventually lost in the presence of many low amplitude SAWs (threshold value $\delta B/B\sim\mathcal{O}(10^{-4})$) \citep{RWhitePoP2010,RWhitePPCF2010}.
Thus, the assessment of the nonlinear RSAE saturation mechanism and amplitude plays a crucial role in evaluating the operation scenario and the EP confinement property.

In general, the channels of SAW nonlinear saturation can be classified into two routes, namely,
the wave-particle nonlinear interaction and wave-wave nonlinear couplings \citep{LChenRMP2016,LChenPoP2013,FZoncaNJP2015,HBerkPoFB1990c}.
The former focuses on the perturbation to the resonant EP phase space distribution function by finite amplitude SAWs \citep{FZoncaNJP2015,HBerkPoFB1990c},
and is widely investigated by numerical simulations, as reviewed in Refs. \citep{LChenRMP2016,PLauberPR2013}.
By contrast, the latter is relatively less explored.
Most of the previous analytical works consider the toroidal Alfv\'en eigenmode (TAE) \citep{CZChengAP1985} as a paradigm case, including the saturation
via ion induced scattering \citep{TSHahmPRL1995,ZQiuNF2019a}, nonlinear modification to the SAW continuum structure \citep{FZoncaPRL1995,LChenPPCF1998},
the spontaneous generation of zero frequency zonal structures (ZFZS) \citep{LChenPRL2012,ZQiuPoP2016,ZQiuNF2017}
as well as geodesic acoustic mode (GAM) \citep{ZQiuPRL2018,ZQiuNF2019b}.
Since the RSAEs are expected to be prevalent in future steady-state burning plasmas, the RSAE saturation via wave-wave nonlinearity deserves special attention.
In particular, the toroidally symmetric zonal field structures \citep{FZoncaJPCS2021}, including the ZFZS, are well known to play important roles in regulating drift wave turbulences \citep{TSHahmPoP1999,ZLinScience1998,LChenPoP2000,PDiamondPPCF2005}
including drift Alfv\'en waves, and thus, leading to cross-scale couplings \citep{FZoncaPPCF2015}
and nonlinear saturation via scattering them to short radial wavelength regime (or shearing in some literatures).
In this work,   spontaneous ZFZS excitation by RSAE modulational instability is analyzed
using nonlinear gyrokinetic theory.

As noted above, the spontaneous excitation of ZFZS by TAE is first discussed in \citep{LChenPRL2012}.
It is shown that under certain conditions, the zonal current (ZC)   is preferentially excited over the electrostatic zonal flow (ZF),
with the branch ratio of ZF/ZC excitation determined by various geometry effects, including the breaking of pure Alfv\'enic state by toroidicity and neoclassical shielding of ZF \citep{MRosenbluthPRL1998}.
In contrast to the TAE case, it is shown in Ref. \citep{ZQiuNF2016} that for beta-induced Alfv\'en eigenmode (BAE),
the excitation of ZF generally dominates, due to the $|k_{\parallel}v_{\rm A}/\omega|\ll1$ ordering.
For the case of RSAE analyzed herein, its frequency is sensitively determined by the value of $q_{\rm min}$ and the underlying values of toroidal/poloidal mode numbers $n$/$m$,
and generally sweeps in-between the typical BAE to TAE frequency ranges.
It is shown that depending on the specific plasma parameters including $k_{\parallel}$, both ZC and ZF generation may dominate,
and the previous conclusions on TAE \citep{LChenPRL2012} and BAE \citep{ZQiuNF2016} can be recovered as limiting cases
of the general nonlinear dispersion relation derived without assuming specific plasma parameters.

We note that, several nonlinear processes, with comparable cross-sections, may be comparably important in saturating AEs, as addressed in Ref. \citep{ZQiuIAEAFEC2018}.
In particular, a channel unique for RSAE saturation is proposed in this work. Due to the ZC  and the associated perturbed poloidal magnetic field generation, the $q$-profile is modulated, which leads to  the modification of the local SAW continuum in the vicinity of $q_{\rm min}$, and consequently RSAE saturation.
The relevance of this channel on RSAE nonlinear saturation is analyzed and evaluated.

This paper is arranged as follows. In section \ref{sec:theoretical model}, the theoretical model is given.
In section \ref{sec:general nonlinear equations}, the general nonlinear equations describing RSAE evolution and ZFZS excitation
are derived. Section \ref{sec:ZFZS spontaneous excitation by RSAE via modulational instability} is devoted to study the linear growth stage of
the modulational instability; and the nonlinear saturation of RSAE via ZFZS scattering is investigated in section \ref{sec:nonlinear RSAE saturation}.
Finally, brief conclusion and discussion are given in section \ref{sec:conclusion and discussion}.

\section{Theoretical model}
\label{sec:theoretical model}

The nonlinear evolution of this system is studied using the standard nonlinear perturbation theory,
considering a shifted circular tokamak equilibrium described by a set of field-aligned flux coordinates $(r, \theta, \varphi)$.
The perturbed fields are represented by two field variables, namely, the electrostatic potential $\delta\phi$ and
the parallel component of vector potential $\delta A_{\parallel}$, while the parallel magnetic field fluctuation $\delta B_{\parallel}$
is suppressed, consistent with $\beta\ll1$ ordering of typical laboratory plasmas.
Here, $\beta$ is the ratio of thermal to magnetic pressures.
For convenience, $\delta A_{\parallel}$ is replaced by $\delta\psi\equiv\omega\delta A_{\parallel}/(ck_{\parallel})$, such that the ideal MHD limit, i.e., vanishing
parallel electric field fluctuation $\delta E_{\parallel}$ corresponds to simply $\delta\phi=\delta\psi$.
In this work, it is assumed that the RSAE is excited by a source outside this nonlinear system, such as EPs,
and the nonlinear coupling is dominated by bulk plasma contribution. For the cases with EPs contributing significantly to the nonlinear ZFZS generation \citep{YTodoNF2010,ABiancalaniJPP2020},
interested readers may refer to Refs. \citep{ZQiuPoP2016,ZQiuNF2017} for more systematic discussions.
To start with, we consider the two-field coupled system which consists of a RSAE (subscript `R') and ZFZS (subscript `Z'),
i.e., $\delta\phi=\delta\phi_{R}+\delta\phi_{Z}$ with $\delta\phi_{R}=\delta\phi_{0}+\delta\phi_{0^{*}}$.
Here, $\delta\phi_{0}$ represents RSAE with positive real frequency and $\delta\phi_{0^{*}}$ represents the counterpart with negative real frequency,
of which there may be a rich spectrum of different radial eigen-states.

Considering the reactor relevant parameter regime with $nq\gg1$, the ballooning mode representation \citep{JConnorPRL1978} for RSAE is adopted,
\begin{eqnarray}
\delta\phi_{0}  & = & A_{0}e^{i\left(n\phi-m_{0}\theta-\omega_{0}t\right)}e^{i\int\hat{k}_{r,0}dr}\sum\limits_{j}e^{-ij\theta}\Phi_{0}\left(x-j\right)+c.c..\nonumber
\end{eqnarray}
Here, $m=m_{0}+j$ with $m_{0}$ being the reference poloidal mode number, $x\equiv nq-m_{0}$,
$\Phi_{0}$ is the parallel mode structure with the typical radial extension comparable to distance between neighboring mode rational surfaces, $A_{0}$ is the mode envelope amplitude
and $\hat{k}_{r,0}$ is the radial envelope wavenumber accounting for the slowly varying radial structures.
Note that, RSAE is typically characterized by one dominant poloidal harmonic, while multiple sub-dominant poloidal harmonics exist due to toroidicity.
Furthermore, $\int|\Phi_{0}|^{2}dx=1$ is used as normalization condition.

Consequently, the ZFZS is expected to have a  fine radial structure in addition to the well-known meso-scale structure \citep{PDiamondPPCF2005}, as a result of the RSAE fine radial mode structure highly localized around $q_{\rm min}$.
For ZFZS dominated by $n=0, m=0$ scalar potential perturbation \citep{LChenPoP2000}, we take
\begin{eqnarray}
\delta\phi_{Z} & = & A_{Z}e^{i\int\hat{k}_{Z}dr-i\omega_{Z}t}\sum\limits_{j}\Phi_{Z}\left(x-j\right)+c.c.,\nonumber
\end{eqnarray}
with $\Phi_{Z}$ accounting for the fine radial structure \citep{ZQiuNF2016} due to nonlinear mode coupling and $A_{Z}\exp{i\int\hat{k}_{Z}dr}$ being the well-known meso-scale structure.
This general representation adopted here can be applied to recover the results obtained from linear growth stage of the modulational instability,
by separating RSAE into pump and upper/lower sidebands, as often used in previous papers \citep{LChenPoP2000};
and is also recovered in section \ref{sec:ZFZS spontaneous excitation by RSAE via modulational instability},
from the derived general nonlinear equations.

The governing equations describing the nonlinear processes can be derived from quasi-neutrality condition
\begin{eqnarray}
\frac{n_{0}e^{2}}{T_{i}}\left(1+\frac{T_{i}}{T_{e}}\right)\delta\phi_{k} & = & \sum\limits_{s}\left\langle qJ_{k}\delta H_{k}\right\rangle_{s},\label{eq:Q.N.}
\end{eqnarray}
and nonlinear gyrokinetic vorticity equation derived from parallel Amp\`ere's law \citep{LChenRMP2016}
\begin{eqnarray}
\frac{c^{2}}{4\pi\omega_{k}^{2}}B\frac{\partial}{\partial l}\frac{k_{\perp}^{2}}{B}\frac{\partial}{\partial l}\delta\psi_k + \frac{e^{2}}{T_{i}}\left\langle \left(1-J_{k}^{2}\right)F_{0}\right\rangle_{s} \delta\phi_{k}-\sum\limits_{s}\left\langle \frac{q}{\omega_{k}}J_{k}\omega_{d}\delta H_{k}\right\rangle\nonumber
\\=-i\frac{c}{B_{0}\omega_{k}}\sum\limits_{\mathbf{k}=\mathbf{k'}+\mathbf{k''}}\mathbf{\hat{b}}\cdot\mathbf{k''}\times\mathbf{k'}\left[\frac{c^{2}}{4\pi}k_{\perp}''^{2}\frac{\partial_{l}\delta\psi_{k'}\partial_{l}\delta\psi_{k''}}{\omega_{k'}\omega_{k''}}\right.\nonumber
\\\left.+\left\langle e\left(J_{k}J_{k'}-J_{k''}\right)\delta L_{k'}\delta H_{k''}\right\rangle \right].\label{eq:vorticity equation}
\end{eqnarray}
Here, $J_{k}\equiv J_{0}(k_{\perp}\rho)$ with
$J_{0}$ being the Bessel function of zero index accounting for finite Larmor radius effects, $\rho=v_{\perp}/\Omega_{c}$ is the Larmor radius with $\Omega_{c}$ being the cyclotron frequency, $F_{0}$ is the
equilibrium particle distribution function, $\omega_{d}=(v_{\perp}^{2}+2v_{\parallel}^{2})(k_{r}\sin\theta+k_{\theta}\cos\theta)/(2\Omega_{c}R)$
is the magnetic drift frequency, $l$ is the length along the equilibrium
magnetic field line, $\left\langle \cdots\right\rangle $ means velocity
space integration, $\sum_{s}$ is the summation of different
particle species with $s=i, e$ representing ion and electron, and $\delta L_{k}\equiv\delta\phi_{k}-k_{\parallel}v_{\parallel}\delta\psi_{k}/\omega_{k}$.
The three terms on the left hand side of Eq. (\ref{eq:vorticity equation}) are, respectively, the field line bending, inertial and curvature coupling terms, dominating the linear SAW physics.
The two terms on the right hand side of Eq. (\ref{eq:vorticity equation})
correspond to Maxwell (MX) and Reynolds stresses (RS) \citep{LChenNF2001} that contribute to nonlinear mode couplings as
MX and RS do not cancel each other, with their contribution dominating in the radially fast varying inertial layer \citep{LChenPoP2000}, and $\sum_{\mathbf{k}=\mathbf{k'}+\mathbf{k''}}$ indicates
the wavenumber and frequency matching condition required for nonlinear mode coupling. $\delta H_{k}$ is the nonadiabatic particle
response, which can be derived from nonlinear gyrokinetic equation \citep{EFriemanPoF1982}:
\begin{eqnarray}
\left(-i\omega_{k}+v_{\parallel}\partial_{l}+i\omega_{d}\right)\delta H_{k} = -i\omega_{k}\frac{q}{T}F_{0}J_{k}\delta L_{k}\nonumber
\\-\frac{c}{B_{0}}\sum\limits_{\mathbf{k}=\mathbf{k'}+\mathbf{k''}}\mathbf{\hat{b}}\cdot\mathbf{k''}\times\mathbf{k'}J_{k'}\delta L_{k'}\delta H_{k''}.\label{eq:gyrokinetic equation}
\end{eqnarray}

For RSAE with $|k_{\parallel}v_{e}|\gg|\omega_k|\gg|k_{\parallel}v_{i}|, |\omega_{d}|$,
the linear ion/electron responses can be derived to the leading order as
$\delta H^{L}_{k, i}=eF_{0}J_k\delta\phi_ k/T_{i}$ and $\delta H^{L}_{k,e}=-eF_{0}\delta\psi_{k}/T_{e}$.
Furthermore, one can have that, to the leading order, ideal MHD constraint is satisfied, i.e., $\delta\phi_{R}=\delta\psi_{R}$,
by substituting the ion/electron responses of RSAE into quasi-neutrality condition.

On the other hand, considering such a nonlinear system dominated by SAW instabilities, we can also use the parallel component of the nonlinear ideal Ohm's law as an alternative to Eq. (\ref{eq:Q.N.}),
\begin{eqnarray}
\delta E_{\parallel,k} & = & -\sum\limits_{\mathbf{k}=\mathbf{k'}+\mathbf{k''}}\mathbf{\hat{b}}\cdot\delta\mathbf{u}_{k'}\times\delta\mathbf{B}_{k''}/c,\label{eq:Ohm's law}
\end{eqnarray}
with $\delta\mathbf{u}$ being the $\mathbf{E}\times\mathbf{B}$ drift velocity.
We note that Eq. (\ref{eq:Ohm's law}) is equivalent to Eq. (\ref{eq:Q.N.}), ignoring the high order $\mathcal{O}(k_{\perp}^{2}\rho_{i}^{2})$ corrections.

\section{General nonlinear equations}
\label{sec:general nonlinear equations}

In this section, the general nonlinear equations describing the self-consistent RSAE evolution are derived, including the generation of ZFZS
and the feedback modulation of RSAE by ZFZS.
Generally speaking, the nonlinear process can be divided into two stages, i.e., linear growth stage and strongly nonlinear stage,
determined by whether the modulation to the pump wave is small.
The governing equations derived in this section, without separating RSAE into pump wave and its sidebands, are general,
and can be used for describing both stages as shown in later sections \citep{ZGuoPRL2009,NChen2021}.

Considering the nonlinear coupling dominated by the radially fast varying inertial region,
one can obtain the equation describing the electrostatic ZF excitation from surface averaged vorticity equation as
\begin{eqnarray}
\omega_{Z}\hat{\chi}_{Z}\delta\phi_{Z} & = & -i\frac{c}{B}k_{\theta}\left(1-\frac{k_{\parallel,0}^{2}v_{\rm A}^{2}}{\omega_{0}^{2}}\right)\left(k_{r,0}-k_{r,0^{*}}\right)\left|\delta\phi_{0}\right|^{2}.\label{eq:ZF}
\end{eqnarray}
Here, $\hat{\chi}_{Z}=\chi_{Z}/(k_{Z}^{2}\rho_{i}^{2})\simeq 1.6q^{2}\epsilon^{-1/2}$ with $\chi_{Z}$ being the well-known neoclassical shielding of ZFZS \citep{MRosenbluthPRL1998}
and $\epsilon\equiv r/R$ being the inverse aspect ratio of the torus. One can note that,
$(k_{r,0}-k_{r,0^{*}})|\delta\Phi_{0}|^{2}\equiv [(\hat{k}_{r,0}-\hat{k}_{r,0^{*}})-i(\partial_{r}\ln\Phi_{0}-\partial_{r}\ln\Phi_{0^{*}})]|\delta\Phi_{0}|^{2}$
being radial modulation with $(\hat{k}_{r,0}-\hat{k}_{r,0^{*}})$ denoting envelope modulation \citep{LChenPRL2012} and $(\partial_{r}\ln\Phi_{0}-\partial_{r}\ln\Phi_{0^{*}})$ denoting
parallel mode structure evolution \citep{ZQiuNF2016}, which gives the fine radial structure of ZFZS.
For RSAE typically dominated by one or two poloidal harmonics, $(\partial_{r}\ln\Phi_{0}-\partial_{r}\ln\Phi_{0^{*}})$ is the dominant term, and determines the zonal structure radial wavenumber $k_{Z}=-i(\partial_{r} \ln\Phi_{0}-\partial_{r}\ln\Phi_{0^{*}})$, as addressed in Ref. \citep{ZQiuNF2017}.

The equation describing the electromagnetic ZC excitation can be derived from Eq. (\ref{eq:Ohm's law}),
considering $k_{\parallel,Z}=0$ and noting $\delta\psi_{Z}\equiv\omega_{0}\delta A_{\parallel,Z}/(ck_{\parallel,0})$ is defined using the frequency and parallel wavenumber of RSAE, as
\begin{eqnarray}
\delta\psi_{Z} & = & -i\frac{c}{B}k_{\theta,0}k_{Z}\frac{1}{\omega_{0}}\left|\delta\phi_{0}\right|^{2}.\label{eq:ZC}
\end{eqnarray}
In deriving Eq.(\ref{eq:ZC}), ideal MHD condition for RSAE ($\delta\phi_{0}=\delta\psi_{0}$) is used,
and $\partial_{r}\ln\delta\psi_{Z}=\partial_{r}\ln|\delta\phi_{0}|^{2}$ is noted.

One can also derive the corresponding equations describing RSAE from Eq. (\ref{eq:Ohm's law}) as
\begin{eqnarray}
\delta\phi_{0}-\delta\psi_{0} & = & -i\frac{c}{B}\frac{k_{Z}k_{\theta,0}}{\omega_{0}}\delta\phi_{0}\left(\delta\phi_{Z}-\delta\psi_{Z}\right),\label{eq:RSAE 1}
\end{eqnarray}
which describes the deviation from ideal MHD constraint due to nonlinear ZFZS modulation.
The other equation for RSAE can be derived from nonlinear vorticity equation as
\begin{eqnarray}
k_{\perp,0}^{2}\left(-\frac{k_{\parallel,0}^{2}v_{\rm A}^{2}}{\omega_{0}^{2}}\delta\psi_{0}+\delta\phi_{0}-\frac{\omega_{G}^{2}}{\omega_{0}^{2}}\delta\phi_{0}\right)\nonumber
\\= -i\frac{c}{B\omega_{0}}k_{Z}k_{\theta,0}\left(k_{Z}^{2}-k_{\theta,0}^{2}\right)\delta\phi_{0}\left(\delta\phi_{Z}-\frac{k_{\parallel,0}^{2}v_{\rm A}^{2}}{\omega_{0}^{2}}\delta\psi_{Z}\right),\label{eq:RSAE 2}
\end{eqnarray}
with the term proportional to $\delta\phi_{Z}$ on the right hand side corresponding to RS contribution and $\delta\psi_{Z}$ term corresponding to MX contribution.
The third term on the left hand side of Eq. (\ref{eq:RSAE 2}) is the SAW continuum upshift due to geodesic curvature induced compression, and $\omega_{G}$ is the frequency of GAM \citep{ZQiuPPCF2009}.

Substituting Eq. (\ref{eq:RSAE 1}) into Eq. (\ref{eq:RSAE 2}), one then obtains the equation describing the modulation of RSAE by ZFZS,
\begin{eqnarray}
k_{\perp,0}^{2}\mathscr{E}_{0}\delta\phi_{0} & = & -i\frac{c}{B\omega_{0}}\left(k_{Z}^{2}-k_{\theta,0}^{2}-\frac{k_{\parallel,0}^{2}v_{\rm A}^{2}}{\omega_{0}^{2}}k_{\perp,0}^{2}\right)
k_{Z}k_{\theta,0}\delta\phi_{0}\left(\delta\phi_{Z}-\alpha\delta\psi_{Z}\right),\label{eq:RSAE 3}
\end{eqnarray}
with $\mathscr{E}_0$ being the RSAE dispersion relation,
and $\alpha\equiv(k_{\parallel,0}^{2}v_{\rm A}^{2}/\omega_{0}^{2})(-2k_{\theta,0}^{2})/(k_{Z}^{2}-k_{\theta,0}^{2}-k_{\parallel,0}^{2}v_{\rm A}^{2}k_{\perp,0}^{2}/\omega_{0}^{2})$.
Eq. (\ref{eq:RSAE 3}) is general, and can be reduced to various limits depending on different plasma parameters, through the
value of $\alpha$ dependence on $k_{\parallel,0}$;
e.g., the mode dynamics described by Eq. (\ref{eq:RSAE 3}) is similar to the TAE case with $(k_{\parallel,0}^{2}v_{\rm A}^{2}/\omega_{0}^{2})\sim\mathcal{O}(1)$,
and $|k_{\parallel,0}|\simeq 1/(2qR)$, and thus $\alpha\simeq 1$. On the other hand, the mode
behavior gets close to a BAE with $k_{\parallel,0}\simeq 0$, and thus $\alpha\simeq 0$.
For simplicity of investigation, the RSAE WKB dispersion relation can be adopted here
as $\mathscr{E}_{0}\simeq(1-k_{\parallel,0}^{2}v_{\rm A}^{2}/\omega_{0}^{2}-\omega_{G}^{2}/\omega_{0}^{2})$,
while the radial global dispersion relation \citep{FZoncaPoP2002} can be applied for more quantitative analysis.

Furthermore, subtracting Eq. (\ref{eq:ZF}) by $\alpha\times$Eq. (\ref{eq:ZC}), one can obtain,
\begin{eqnarray}
\delta\phi_{Z}-\alpha\delta\psi_{Z} & = & -i\frac{c}{B}k_{Z}k_{\theta,0}
\left[\frac{1-k_{\parallel,0}^{2}v_{\rm A}^{2}/\omega_{0}^{2}}{\omega_{Z}\hat{\chi}_{Z}k_{Z}}\left(k_{r,0}-k_{r,0^{*}}\right)+\frac{\alpha}{\omega_{0}}\right]\left|\delta\phi_{0}\right|^{2},
\label{eq:ZFZS}
\end{eqnarray}
which can be substituted into Eq. (\ref{eq:RSAE 3}), and yield the general equation describing the self-modulation of RSAE, as
\begin{eqnarray}
k_{\perp,0}^{2}\mathscr{E}_{0}\delta\phi_{0} & = & -\left(\frac{c}{B}k_{Z}k_{\theta,0}\right)^{2}
\frac{1}{\omega_{0}}\left(k_{Z}^{2}-k_{\theta,0}^{2}-\frac{k_{\parallel0}^{2}v_{\rm A}^{2}}{\omega_{0}^{2}}k_{\perp0}^{2}\right)\nonumber
\\&\times&\left[\frac{1-k_{\parallel,0}^{2}v_{\rm A}^{2}/\omega_{0}^{2}}{\omega_{Z}\hat{\chi}_{Z}k_{Z}}\left(k_{r,0}-k_{r,0^{*}}\right)+\frac{\alpha}{\omega_{0}}\right]\delta\phi_{0}\left|\delta\phi_{0}\right|^{2}.
\label{eq:RSAE final}
\end{eqnarray}
Both ZF and ZC generation by RSAE are systematically accounted for in Eq. (\ref{eq:RSAE final}) on the same footing,
with the first term in square brackets corresponds to the ZF generation while the second term corresponds to the ZC.
On the one hand, which one is preferentially excited is shown in the section \ref{sec:ZFZS spontaneous excitation by RSAE via modulational instability}.
On the other hand, ZFZS generation can lead to RSAE saturation by scattering it to linearly stable radial eigen-states.
The nonlinear saturation level, can be determined by self-consistently solving Eq. (\ref{eq:RSAE final}) as a nonlinear Schrodinger equation \citep{NChen2021},
while a rough estimation is given in, e.g., Ref. \citep{ZQiuPRL2018}, by separating the AE into a pump and its sidebands, and deriving the fixed point solution of the coupled nonlinear equations.
Since RSAE linear properties is very sensitive to $q$-profile, the modulation of $q$-profile caused by the nonlinearly generated ZC
may have a great impact on RSAE nonlinear saturation. This is discussed in the section \ref{sec:nonlinear RSAE saturation}.

\section{ZFZS spontaneous excitation by RSAE via modulational instability}
\label{sec:ZFZS spontaneous excitation by RSAE via modulational instability}

To investigate the linear growth stage of the modulational instability, we follow the analysis of Ref. \citep{LChenPRL2012}, and consider the fluctuation consists of a constant-amplitude pump wave
$\Omega_{P}\equiv\Omega_{P}(\omega_{P}, \mathbf{k}_{P})$ and its upper and lower sidebands $\Omega_{\pm}\equiv\Omega_{\pm}(\omega_{\pm}, \mathbf{k}_{\pm})$
due to the modulation of the ZFZS $\Omega_{Z}\equiv\Omega_{Z}(\omega_{Z}, \mathbf{k}_{Z})$ \citep{LChenPoP2000}.
Here, subscripts `$P$', `$+$' and `$-$' denote RSAE pump, upper and lower sidebands respectively, with $\delta\phi_{0}=\delta\phi_{P}+\delta\phi_{+}$,
$\delta\phi_{0^{*}}=\delta\phi_{P^{*}}+\delta\phi_{-}$, and
\begin{eqnarray}
\delta\phi_{P} & = & A_{P}e^{i\left(n\phi-m_{0}\theta-\omega_{P}t\right)}\sum\limits_{j}e^{-ij\theta}\Phi_{P}\left(x-j\right)+c.c.,\nonumber\\
\delta\phi_{\pm} & = & A_{\pm}e^{\pm i\left(n\phi-m_{0}\theta-\omega_{P}t\right)}e^{i\left(\int\hat{k_{Z}}dr-\omega_{Z}t\right)}
\sum\limits_{j}e^{\mp ij\theta}\left\{ \begin{array}{c}
\Phi_{P}\left(x-j\right)\\
\Phi_{P^{*}}\left(x-j\right)
\end{array}\right\} +c.c..\nonumber
\end{eqnarray}
Then the general nonlinear Eqs. (\ref{eq:ZF})-(\ref{eq:RSAE 2}) can be reduced to  equations describing    $\delta\phi_{Z}$, $\delta\phi_{+}$ and $\delta\phi_{-}$ generation by the fixed amplitude pump RSAE, while the feedback of ZFZS and RSAE sidebands to the pump wave is neglected, focusing on the initial stage of the nonlinear process.
Considering the frequency/wave number matching condition ($\omega_{\pm}=\pm\omega_{P}+\omega_{Z}$, $\mathbf{k}_{\pm}=\pm\mathbf{k}_{P}+\mathbf{k}_{Z}$) imbedded in the above expressions,
Eq. (\ref{eq:ZF}) can be reduced to
\begin{eqnarray}
\omega_{Z}\hat{\chi}_{Z}\delta\phi_{Z} & = & -i\frac{c}{B}k_{Z}k_{\theta,P}\left(1-\frac{k_{\parallel,P}^{2}v_{\rm A}^{2}}{\omega_{P}^{2}}\right)
\left(\delta\phi_{+}\delta\phi_{P^{*}}-\delta\phi_{-}\delta\phi_{P}\right),\label{eq:ZF sideband}
\end{eqnarray}
with $(1-k^{2}_{\parallel,P}v^{2}_{\rm A}/\omega^{2}_{P})$ representing the competition of Reynolds and Maxwell stresses to break the pure Alfv\'enic state \citep{LChenPoP2013,LChenPRL2012}.
On the other hand, Eq. (\ref{eq:ZC}) describing ZC excitation can be reduced to
\begin{eqnarray}
\delta\psi_{Z} & = & -i\frac{c}{B}k_{\theta,P}k_{Z}\frac{1}{\omega_{P}}\left(\delta\phi_{+}\delta\phi_{P^{*}}+\delta\phi_{-}\delta\phi_{P}\right).\label{eq:ZC sideband}
\end{eqnarray}
The nonlinear equation describing RSAE sidebands generation through the ZFZS modulation to pump RSAE can be derived from Eq. (\ref{eq:RSAE 3}) as
\begin{eqnarray}
k_{\perp,\pm}^{2}\mathscr{E}_{\pm}\delta\phi_{\pm} & = & -i\frac{c}{B\omega_{\pm}}\left(k_{Z}^{2}-k_{\theta,P}^{2}-\frac{k_{\parallel,P}^{2}v_{\rm A}^{2}}{\omega_{P}^{2}}k_{\perp,\pm}^{2}\right)\nonumber
\\&\times&k_{Z}k_{\theta,P}\left\{ \begin{array}{c}
\delta\phi_{P}\\
\delta\phi_{P^{*}}
\end{array}\right\} \left(\delta\phi_{Z}-\alpha\delta\psi_{Z}\right).\label{eq:RSAE sideband}
\end{eqnarray}
Eqs. (\ref{eq:ZF sideband})-(\ref{eq:RSAE sideband}), are equivalent to Eqs. (34)-(36) for TAE cases as derived in Ref. \citep{ZQiuNF2017},
with the coefficient $\alpha$ generalized to include a broader parameter regime ($\alpha\simeq 1$ for TAE as discussed in Ref. \citep{ZQiuNF2017}).
Note that $k_{\perp,\pm}^{2}=k_{\perp,P}^{2}+k_{Z}^{2}$ and $k_{\perp,+}^{2}$ is used in the derivation later.
Similarly, subtracting Eq. (\ref{eq:ZF sideband}) by $\alpha\times$Eq. (\ref{eq:ZC sideband}), one can obtain
\begin{eqnarray}
\delta\phi_{Z}-\alpha\delta\psi_{Z} & = & i\frac{c}{B}k_{Z}k_{\theta,P}
\left[\left(\frac{1-k_{\parallel,P}^{2}v_{\rm A}^{2}/\omega_{P}^{2}}{\omega_{Z}\hat{\chi}_{Z}}+\frac{\alpha}{\omega_{P}}\right)\delta\phi_{+}\delta\phi_{P^{*}}\right.\nonumber
\\&&\left.-\left(\frac{1-k_{\parallel,P}^{2}v_{\rm A}^{2}/\omega_{P}^{2}}{\omega_{Z}\hat{\chi}_{Z}}-\frac{\alpha}{\omega_{P}}\right)\delta\phi_{-}\delta\phi_{P}\right].
\label{eq:ZFZS sideband}
\end{eqnarray}
The modulational instability dispersion relation can then be derived by substituting Eq. (\ref{eq:RSAE sideband}) into (\ref{eq:ZFZS sideband}), as
\begin{eqnarray}
1 & = & -\hat{F}\left|\delta\phi_{P}\right|^{2}\left[\left(\frac{1-k_{\parallel,P}^{2}v_{\rm A}^{2}/\omega_{P}^{2}}{\omega_{Z}\hat{\chi}_{Z}}+\frac{\alpha}{\omega_{P}}\right)\frac{1}{\mathscr{E}_{+}}\right.\nonumber
\\&&-\left.\left(\frac{1-k_{\parallel,P}^{2}v_{\rm A}^{2}/\omega_{P}^{2}}{\omega_{Z}\hat{\chi}_{Z}}-\frac{\alpha}{\omega_{P}}\right)\frac{1}{\mathscr{E}_{-}}\right],\label{eq:linear growth stage D.R.}
\end{eqnarray}
with $\hat{F}=(ck_{Z}k_{\theta,P}/B)^{2}(-k_{Z}^{2}+k_{\theta,P}^{2}+k_{\parallel,P}^{2}v_{\rm A}^{2}k_{\perp,+}^{2}/\omega_{P}^{2})/(\omega_{P}k_{\perp,+}^{2})$
being a nonlinear coupling coefficient.
Furthermore, considering that RSAE sidebands still obey the dispersion relation of RSAE by $\mathscr{E}_{\pm}=\mathscr{E}_{0}(\omega_Z\pm\omega_P, k_{Z})$, one can expand $\mathscr{E}_{\pm}$ along
the RSAE characteristics, as $\mathscr{E}_{\pm}\simeq(\partial\mathscr{E}_{0}/\partial\omega_{0})(\pm\omega_{Z}-\Delta)$ with
$\Delta\equiv -k_{Z}^{2}(\partial^{2}\mathscr{E}_{0}/\partial k_{r}^{2})/(2\partial\mathscr{E}_{0}/\partial\omega_{0})$ being the frequency mismatch,
describing the frequency shift of RSAE sidebands from the pump RSAE dispersion relation due to the ZFZS modulation.
Denoting $\gamma\equiv-i\omega_{Z}$
and noting $\omega_{\pm}\simeq\pm\omega_{P}$, the modulational instability dispersion relation can be shown as
\begin{eqnarray}
\gamma^{2} & = & -\Delta^{2}+\frac{2\hat{F}\left|\delta\phi_{P}\right|^{2}}{\partial\mathscr{E}_{0}/\partial\omega_{0}}
\left(\frac{1-k_{\parallel,P}^{2}v_{\rm A}^{2}/\omega_{0}^{2}}{\hat{\chi}_{Z}}+\frac{\alpha}{\omega_{P}}\Delta\right).\label{eq:growth rate D.R.}
\end{eqnarray}
Here, the first term on the right side of Eq. (\ref{eq:growth rate D.R.}) is the threshold condition due to frequency mismatch and the second term represents the nonlinear drive.
Thus, the ZFZS can be spontaneously excited when the nonlinear drive overcomes the threshold condition due to frequency mismatch,
as the RSAE amplitude is large enough, or the nonlinear coupling is strong enough, as we address in the following discussion.
Furthermore, the first term in brackets corresponds to ZF contribution to the nonlinear coupling,  and the second term  corresponds to ZC.
For ZF contribution, there are two restrictions due to, first, the partial cancelation of RS and MX
with the `residual' drive due to deviation from ideal MHD limit due to plasma nonuniformity (reversed $q$-profile here) that breaks the pure Alfv\'enic state \citep{FZoncaPoP2002,LChenPoP2013},
and second, the neoclassical shielding of ZF as shown by the $\hat{\chi}_{Z}$ in the denominator,
with $\hat{\chi}_{Z}\sim q^{2}/\epsilon$, which is typically much larger than unity \citep{LChenRMP2016,MRosenbluthPRL1998}.

For ZC contribution, there are also two important factors that crucially determine the nonlinear process, with the first being the sign of $\Delta$, which is typically determined by specific plasma parameters.
The ZC term is a driving term for $\Delta>0$; while for $\Delta<0$, the ZC term becomes a damping term for this nonlinear process;
and then the condition for modulational instability becomes very stringent, with additional requirement for the ZF term being dominant over the ZC term,
which is not easy to satisfy due to the two restrictions as addressed in the paragraph above.
The other important factor is the value of $\alpha$. As noted before, if RSAE localizes near the rational surface, $\alpha\simeq 0$ because of $k_{\parallel}\simeq 0$,
meaning ZC generation is very weak and is similar to BAE case investigated in Ref. \citep{ZQiuNF2016}; on the other hand, if RSAE localizes in the middle of two neighboring rational surfaces, $\alpha\simeq 1$,
and this case is similar to the TAE case \citep{LChenPRL2012} with ZC generation preferred.
So for RSAE, with $k_{\parallel}$ and its frequency determined by $q_{\rm min}$ and the corresponding mode numbers, both ZF and ZC generation could be dominant,
depending on the specific plasma parameters, and should be investigated case by case.

\section{Nonlinear RSAE saturation}
\label{sec:nonlinear RSAE saturation}

In strongly nonlinear stage, the feedback of ZFZS and RSAE sidebands to the pump wave can not be neglected
anymore, as the sideband amplitudes become comparable to that of the pump wave.
It is indicated in Eq. (\ref{eq:RSAE final}) that ZS can play self-regulatory roles on RSAE nonlinear evolution
by scattering RSAE into short radial wavelength stable domains, which may lead to RSAE saturation.
The saturation level, can be derived from the coupled pump wave and sidebands equations, as is shown in, e.g., Ref. \citep{FZoncaEPL2008}.
In addition, another related channel unique for RSAE nonlinear saturation may exist, due to the modulation to SAW continuum \citep{FZoncaPRL1995,LChenPPCF1998}
by the nonlinearly generated ZS (among which ZC playing a dominant role), considering the sensitive dependence of RSAE on SAW continuum accumulation point (for a visualization of RSAE physics dependence on $q_{\rm min}$, interested readers may refer to Ref. \citep{TWangNF2020} and Fig. 3 therein).
The nonlinearly generated ZC is a toroidal current sharply localized around $q_{\rm min}$, which can generate a perturbed poloidal magnetic field
and further modulate $q$-profile and thus the SAW continuum near $q_{\rm min}$. Thus, one can reasonably speculate that ZC plays an important role in RSAE saturation
by modifying the equilibrium continuum, similar to the mechanism investigated in Ref. \citep{FZoncaPRL1995} and Ref. \citep{LChenPPCF1998} for TAE.

Here, in consistency with the above speculation on the crucial role played by ZC, we consider a simplified case that ZC generation is dominant.
This assumption is natural for scenarios with $(k_{\parallel,0}^{2}v_{\rm A}^{2}/\omega_{0}^{2})\sim\mathcal{O}(1)$,
however, may also be important for other parameter regimes for the reasons addressed above.
Thus, Eq. (\ref{eq:RSAE final}) can be simplified as
\begin{eqnarray}
\mathscr{E}_{0}\delta\phi_{0} & = & 2\left(\frac{c}{B\omega_{0}}k_{Z}k_{\theta,0}\right)^{2}\delta\phi_{0}\left|\delta\phi_{0}\right|^{2}.\label{eq:RSAE ZC}
\end{eqnarray}
Taking a two scale analysis by assuming $\omega_{R}=\omega_{0}+i\partial_{\tau}$, and expanding $\mathscr{E}_{0}\simeq(\partial\mathscr{E}_{0}/\partial\omega_{0})(i\partial_{\tau}-i\gamma^{L}_{R}-\Delta)$ with
$\Delta$ being the nonlinearity induced frequency shift and $\gamma^{L}_{R}$ is the linear RSAE growth rate \footnote{noting the linear RSAE growth/damping rate here, which is not included in Eq. (\ref{eq:growth rate D.R.}) assuming RSAE sidebands damping rates are typically smaller than frequency mismatch}, Eq. (\ref{eq:RSAE ZC}) becomes
\begin{eqnarray}
\left[i\partial_{\tau}-i\gamma^{L}_{R}-\Delta-2\left(\frac{c}{B\omega_{0}}k_{Z}k_{\theta,0}\right)^{2}\frac{\left|\delta\phi_{0}\right|^{2}}{\partial\mathscr{E}_{0}/\partial\omega_{0}}\right]\delta\phi_{0} & = & 0.\label{eq:RSAE saturation}
\end{eqnarray}
Eq. (\ref{eq:RSAE saturation}) describes the RSAE nonlinear evolution due to, scattering to different radial eigen-state (denoted by $\Delta$) with different linear growth/damping rates ($\gamma^{L}_{R}$) and nonlinear self-modulation by ZC generation,
and the saturation level can be estimated by balancing the frequency shift (Max ($|\gamma^{L}_{R}|$, $|\Delta|$)) and the nonlinear RSAE modulation by ZC,
then one has
\begin{eqnarray}
\left|\delta\phi_{0}\right|^{2} & = & \frac{\partial^{2}\mathscr{E}_{0}}{\partial k_{r}^{2}}\left(\frac{B\omega_{0}}{2ck_{\theta,0}}\right)^{2}.\label{eq:RSAE saturation level}
\end{eqnarray}
Eq. (\ref{eq:RSAE saturation level}) describes the RSAE saturation level due to scattering by self-generated ZC to neighboring (linearly stabler) radial eigen-states, assuming $\Delta\gg\gamma^{L}_{R}$.
Cases with $\Delta\ll\gamma^{L}_{R}$ can be evaluated similarly.
The RSAE saturation level can be estimated by substituting typical parameters into the expression. Furthermore, Eq. (\ref{eq:RSAE saturation level}) can be substituted into Eq. (\ref{eq:ZC}), and the saturation level of the perturbed poloidal magnetic field $\delta B_{\theta}$  can be estimated as
\begin{eqnarray}
\delta B_{\theta} & \sim & \frac{B_{0}k_{\parallel,0}k_{Z}^{2}}{4k_{\theta,0}}\left.\frac{\partial^{2}\mathscr{E}_{0}}{\partial k_{r}^{2}}\right|_{k_{r}=0},\label{eq:B modulation}
\end{eqnarray}
resulting in a modulation of local $q_{\rm min}$ by
\begin{eqnarray}
\delta q & \sim & -q_{\rm min}\frac{B_{0}k_{\parallel,0}k_{Z}^{2}}{4B_{\theta}k_{\theta,0}}\left.\frac{\partial^{2}\mathscr{E}_{0}}{\partial k_{r}^{2}}\right|_{k_{r}=0}.\label{eq:q modulation}
\end{eqnarray}
In deriving Eqs. (\ref{eq:B modulation}) and (\ref{eq:q modulation}), we have noted $\delta\psi=\omega\delta A_{\parallel}/(ck_{\parallel})$ and $\mathbf{\delta B}=\nabla\times\delta A_{\parallel}\mathbf{b}$.  Typical parameters can be adopted, i.e.,   $q_{\rm min}\sim\mathcal{O}(1)$, $B_{0}/B_{\theta}\sim qR/r_{0}$,
$k_{\parallel}\sim 1/(qR)$, $k_{\theta,0}\rho_{d}\sim\mathcal{O}(1)$ with $\rho_{d}$ being the drift orbit radius and
\begin{eqnarray}
k_{Z}^{2}\left.\frac{\partial^{2}\mathscr{E}_{0}}{\partial k_{r}^{2}}\right|_{k_{r}=0} \sim \frac{4\Delta}{\omega},\label{eq:ordering}
\end{eqnarray}
which can be reasonably assumed as $\sim \mathcal{O}(0.1)$.
Thus, one can roughly estimate that $\delta q/q\sim\mathcal{O}(10^{-3})$.
Noting that the modification to local RSAE continuum frequency is $\sim O(n\delta qv_{\rm A}/(qR))$, and that for reactor burning plasma with $\rho_{d}/a\sim O(10^{-2})$, most unstable RSAEs are characterized by $n\gtrsim O(10)$ \citep{TWangPoP2018}, the modification to local SAW
continuum is comparable to the RSAE linear growth rate $\gamma^{L}_{R}$
or frequency differences between different radial eigen-states ($\sim \Delta$). Thus, one expects that the ZC induced SAW continuum modification in the vicinity of $q_{\rm min}$, plays an important role in RSAE nonlinear saturation, though the self-consistent study analogous to Refs. \citep{FZoncaPRL1995,LChenPPCF1998} is not presented. The systematic investigation of RSAE saturation due to nonlinear modification of SAW continuum and the resulting enhanced continuum damping, will be presented in a separate publication.

\section{Conclusion and Discussion}
\label{sec:conclusion and discussion}

The general equations describing RSAE self-modulation via nonlinear excitation of ZFZS are derived using gyrokinetic theory, which is then applied to study the spontaneous ZFZS excitation via modulational instability as well as RSAE nonlinear saturation due to scattering to stabler radial eigen-states. It is found that, both  ZF and ZC  can be dominant in the spontaneous excitation by RSAE,
depending on the specific plasma parameters, especially $q_{\rm min}$ that determines the RSAE parallel wavenumber and frequency. The obtained general modulational instability dispersion relation for ZFZS excitation by RSAE, Eq. (\ref{eq:growth rate D.R.}), can recover the results of TAE \citep{LChenPRL2012}
and BAE \citep{ZQiuNF2016} in the proper limits, i.e., by taking $|k_{\parallel}v_{\rm A}/\omega|\rightarrow 1$ and $0$, respectively. The properties of  ZFZS generation by RSAE, noting that  the typical RSAE parallel wavenumber and frequency are in between those of TAE and BAE, can be understood based on the knowledge obtained from TAE \citep{LChenPRL2012} and BAE \citep{ZQiuNF2016}.

An interesting step forward is, the saturation level of RSAE is qualitatively estimated by balancing the nonlinear scattering by ZFZS  with the frequency differences between different radial eigen-states ($\sim \Delta$), assuming ZC playing a dominant role in the RSAE scattering. The corresponding ZC saturation level as well as the modification to local $q_{\rm min}$, are also estimated. It is found that, the resulting modification to local SAW continuum accumulational point frequency,  can be at least, comparable to the RSAE linear growth rate or frequency mismatch between different radial eigen-states for burning plasma scenarios with most unstable  RSAEs characterized by  $n\gtrsim O(10)$ \citep{TWangPoP2018}. Thus, the modification of local SAW continuum by ZC is expected to contribute significantly to RSAE saturation \citep{FZoncaPRL1995,LChenPPCF1998}.

The above estimation based on  Eq. (\ref{eq:RSAE ZC}), assumes dominant ZC generation by taking $|k_{\parallel}v_{\rm A}/\omega|\sim 1$, is valid for other parameter regimes due to the weak coupling coefficients of ZF generation, except for the cases that $q_{\rm min}$ is localized very close to a low-order rational surface, such that the RSAE properties is close to BAE with predominantly ZF generation \citep{ZQiuNF2016}. The logic underlying the reasoning presented in section \ref{sec:nonlinear RSAE saturation} is that, the RSAE and ZFZS saturation levels are estimated without accounting for the modification to SAW continuum, which is then used to estimate the modification to SAW continuum by the saturated ZC, and is found that the modification to SAW continuum could be comparable to or even more important than the ZFZS shearing. Thus, the obtained results indicate that, the modification to SAW continuum, will start to influence the RSAE nonlinear evolution, before it saturates due to self-modulation via ZFZS generation.
Our work thus indicates that multiple processes may contribute comparably to the RSAE saturation, and should be accounted for on the same footing, based on the solid understanding of each individual process, to properly assess the saturation and thus EP transport by RSAE. This is of particular importance since RSAEs are expected to be strongly excited by core-localized fusion alphas in future reactors characterized by the advanced reversed shear scenarios.

\section*{Acknowledgements}

This work is supported by the National Key R\&D Program of China under Grant No. 2017YFE0301900 and
the National Science Foundation of China under grant No. 11875233.
The authors acknowledge Prof. Liu Chen (ZJU\&UCI) and Dr. Fulvio Zonca (ENEA, Italy) for fruitful discussions.

\section*{Data Availability}
The data that support the findings of this study are available from the corresponding author upon reasonable request.

\section*{}

\bibliographystyle{jpp}
\bibliography{refer}

\begin{thebibliography}{51}
\expandafter\ifx\csname natexlab\endcsname\relax\def\natexlab#1{#1}\fi
\def\au#1{#1} \def\ed#1{#1} \def\yr#1{#1}\def\at#1{#1}\def\jt#1{\textit{#1}}
  \def\bt#1{#1}\def\bvol#1{\textbf{#1}} \def\vol#1{#1} \def\pg#1{#1}
  \def\publ#1{#1}\def\arxiv#1{#1}\def\org#1{#1}\def\st#1{\textit{#1}}

\bibitem[Berk \& Breizman(1990)]{HBerkPoFB1990c}
{\sc \au{Berk, H.~L.} \& \au{Breizman, B.~N.}} \yr{1990}  \at{Saturation of a
  single mode driven by an energetic injected beam. iii. alfv\'en wave
  problem}.  \jt{Physics of Fluids B}  \bvol{2}~(9),  \pg{2246}.

\bibitem[Biancalani {\em et~al.\/}(2020)Biancalani, Bottino, Lauber, Mishchenko
  \& Vannini]{ABiancalaniJPP2020}
{\sc \au{Biancalani, A.}, \au{Bottino, A.}, \au{Lauber, P.}, \au{Mishchenko,
  A.} \& \au{Vannini, F.}} \yr{2020}  \at{Effect of the electron redistribution
  on the nonlinear saturation of alfv\'en eigenmodes and the excitation of
  zonal flows}.  \jt{Journal of Plasma Physics}  \bvol{86}~(3),
  \pg{825860301}.

\bibitem[Chen(1999)]{LChenJGR1999}
{\sc \au{Chen, L.}} \yr{1999}  \at{Theory of plasma transport induced by
  low-frequency hydromagnetic waves}.  \jt{Journal of Geophysical Research:
  Space Physics}  \bvol{104}~(A2),  \pg{2421}.

\bibitem[Chen {\em et~al.\/}(2000)Chen, Lin \& White]{LChenPoP2000}
{\sc \au{Chen, L.}, \au{Lin, Z.} \& \au{White, R.}} \yr{2000}  \at{Excitation
  of zonal flow by drift waves in toroidal plasmas}.  \jt{Physics of Plasmas}
  \bvol{7}~(8),  \pg{3129--3132}.

\bibitem[Chen {\em et~al.\/}(2001)Chen, Lin, White \& Zonca]{LChenNF2001}
{\sc \au{Chen, L.}, \au{Lin, Z.}, \au{White, R.~B.} \& \au{Zonca, F.}}
  \yr{2001}  \at{Non-linear zonal dynamics of drift and drift-alfv{\'e}n
  turbulence in tokamak plasmas}.  \jt{Nuclear fusion}  \bvol{41}~(6),
  \pg{747}.

\bibitem[Chen \& Zonca(2012)]{LChenPRL2012}
{\sc \au{Chen, L.} \& \au{Zonca, F.}} \yr{2012}  \at{Nonlinear excitations of
  zonal structures by toroidal alfv\'en eigenmodes}.  \jt{Phys. Rev. Lett.}
  \bvol{109},  \pg{145002}.

\bibitem[Chen \& Zonca(2013)]{LChenPoP2013}
{\sc \au{Chen, L.} \& \au{Zonca, F.}} \yr{2013}  \at{On nonlinear physics of
  shear alfv\'en waves}.  \jt{Physics of Plasmas}  \bvol{20}~(5).

\bibitem[Chen \& Zonca(2016)]{LChenRMP2016}
{\sc \au{Chen, L.} \& \au{Zonca, F.}} \yr{2016}  \at{Alfv\'en waves and
  energetic particles}.  \jt{Review of Modern Physics}  \bvol{88}~(1),
  \pg{015008}.

\bibitem[Chen {\em et~al.\/}(1998)Chen, Zonca, Santoro \& Hu]{LChenPPCF1998}
{\sc \au{Chen, L.}, \au{Zonca, F.}, \au{Santoro, R.} \& \au{Hu, G.}} \yr{1998}
  \at{Nonlinear dynamics of alfv{\'e}n eigenmodes in toroidal plasmas}.
  \jt{Plasma physics and controlled fusion}  \bvol{40}~(11),  \pg{1823}.

\bibitem[Chen {\em et~al.\/}(2021)Chen, Wei, Wei \& Qiu]{NChen2021}
{\sc \au{Chen, N.}, \au{Wei, S.}, \au{Wei, G.} \& \au{Qiu, Z.}} \yr{2021}
  \at{Soliton generation and drift wave turbulence spreading via geodesic
  acoustic mode excitation}.  \jt{submitted to Plasma Physics and Controlled
  Fusion} .

\bibitem[Chen {\em et~al.\/}(2014)Chen, Yu, Liu, Ding, Xie, Zhu, Yu, Ji, Li,
  Li, Yu, Shi, Song, Cao, Song, Dong, Zhong, Jiang, Cui, Huang, Zhou, Dong, Xu,
  Xia, Yan, Yang, Duan \& the HL-2A~Team]{WChenNF2014}
{\sc \au{Chen, W.}, \au{Yu, L.}, \au{Liu, Y.}, \au{Ding, X.}, \au{Xie, H.},
  \au{Zhu, J.}, \au{Yu, L.}, \au{Ji, X.}, \au{Li, J.}, \au{Li, Y.}, \au{Yu,
  D.}, \au{Shi, Z.}, \au{Song, X.}, \au{Cao, J.}, \au{Song, S.}, \au{Dong, Y.},
  \au{Zhong, W.}, \au{Jiang, M.}, \au{Cui, Z.}, \au{Huang, Y.}, \au{Zhou, Y.},
  \au{Dong, J.}, \au{Xu, M.}, \au{Xia, F.}, \au{Yan, L.}, \au{Yang, Q.},
  \au{Duan, X.} \& \au{the HL-2A~Team}} \yr{2014}  \at{Destabilization of
  reversed shear alfv\'en eigenmodes driven by energetic ions during nbi in
  hl-2a plasmas with $q_{\rm min} \sim 1$}.  \jt{Nuclear Fusion}
  \bvol{54}~(10),  \pg{104002}.

\bibitem[Cheng {\em et~al.\/}(1985)Cheng, Chen \& Chance]{CZChengAP1985}
{\sc \au{Cheng, C.}, \au{Chen, L.} \& \au{Chance, M.}} \yr{1985}  \at{High-n
  ideal and resistive shear alfv\'en waves in tokamaks}.  \jt{Ann. Phys.}
  \bvol{161},  \pg{21}.

\bibitem[Connor {\em et~al.\/}(1978)Connor, Hastie \& Taylor]{JConnorPRL1978}
{\sc \au{Connor, J.}, \au{Hastie, R.} \& \au{Taylor, J.}} \yr{1978}  \at{Shear,
  periodicity, and plasma ballooning modes}.  \jt{Phys. Rev. Lett.}
  \bvol{40}~(6),  \pg{396}.

\bibitem[Diamond {\em et~al.\/}(2005)Diamond, Itoh, Itoh \&
  Hahm]{PDiamondPPCF2005}
{\sc \au{Diamond, P.~H.}, \au{Itoh, S.-I.}, \au{Itoh, K.} \& \au{Hahm, T.~S.}}
  \yr{2005}  \at{Zonal flows in plasma: a review}.  \jt{Plasma Physics and
  Controlled Fusion}  \bvol{47}~(5),  \pg{R35}.

\bibitem[Ding {\em et~al.\/}(2015)Ding, Pitts, Borodin, Carpentier, Ding, Gong,
  Guo, Kirschner, Kocan, Li, Luo, Mao, Qian, Stangeby, Wampler, Wang \&
  Wang]{RDingNF2015}
{\sc \au{Ding, R.}, \au{Pitts, R.}, \au{Borodin, D.}, \au{Carpentier, S.},
  \au{Ding, F.}, \au{Gong, X.}, \au{Guo, H.}, \au{Kirschner, A.}, \au{Kocan,
  M.}, \au{Li, J.}, \au{Luo, G.-N.}, \au{Mao, H.}, \au{Qian, J.}, \au{Stangeby,
  P.}, \au{Wampler, W.}, \au{Wang, H.} \& \au{Wang, W.}} \yr{2015}
  \at{Material migration studies with an iter first wall panel proxy on east}.
  \jt{Nuclear Fusion}  \bvol{55}~(2),  \pg{023013}.

\bibitem[Fasoli {\em et~al.\/}(2007)Fasoli, Gormenzano, Berk, Breizman,
  Briguglio, Darrow, Gorelenkov, Heidbrink, Jaun, Konovalov, Nazikian,
  Noterdaeme, Sharapov, Shinohara, Testa, Tobita, Todo, Vlad \&
  Zonca]{AFasoliNF2007}
{\sc \au{Fasoli, A.}, \au{Gormenzano, C.}, \au{Berk, H.}, \au{Breizman, B.},
  \au{Briguglio, S.}, \au{Darrow, D.}, \au{Gorelenkov, N.}, \au{Heidbrink, W.},
  \au{Jaun, A.}, \au{Konovalov, S.}, \au{Nazikian, R.}, \au{Noterdaeme, J.-M.},
  \au{Sharapov, S.}, \au{Shinohara, K.}, \au{Testa, D.}, \au{Tobita, K.},
  \au{Todo, Y.}, \au{Vlad, G.} \& \au{Zonca, F.}} \yr{2007}  \at{Chapter 5:
  Physics of energetic ions}.  \jt{Nuclear Fusion}  \bvol{47}~(6),  \pg{S264}.

\bibitem[Fasoli {\em et~al.\/}(2002)Fasoli, Testa, Sharapov, Berk, Breizman,
  Gondhalekar, Heeter, Mantsinen \& contributors to~the
  EFDA-JET~Workprogramme]{AFasoliPPCF2002}
{\sc \au{Fasoli, A.}, \au{Testa, D.}, \au{Sharapov, S.}, \au{Berk, H.~L.},
  \au{Breizman, B.}, \au{Gondhalekar, A.}, \au{Heeter, R.~F.}, \au{Mantsinen,
  M.} \& \au{contributors to~the EFDA-JET~Workprogramme}} \yr{2002}  \at{{MHD}
  spectroscopy}.  \jt{Plasma Physics and Controlled Fusion}  \bvol{44}~(12B),
  \pg{B159--B172}.

\bibitem[Frieman \& Chen(1982)]{EFriemanPoF1982}
{\sc \au{Frieman, E.~A.} \& \au{Chen, L.}} \yr{1982}  \at{Nonlinear gyrokinetic
  equations for low-frequency electromagnetic waves in general plasma
  equilibria}.  \jt{Physics of Fluids}  \bvol{25}~(3),  \pg{502--508}.

\bibitem[Gormezano {\em et~al.\/}(2007)Gormezano, Sips, Luce, Ide, Becoulet,
  Litaudon, Isayama, Hobirk, Wade, Oikawa, Prater, Zvonkov, Lloyd, Suzuki,
  Barbato, Bonoli, Phillips, Vdovin, Joffrin, Casper, Ferron, Mazon, Moreau,
  Bundy, Kessel, Fukuyama, Hayashi, Imbeaux, Murakami, Polevoi \&
  John]{CGormezanoNF2007}
{\sc \au{Gormezano, C.}, \au{Sips, A.}, \au{Luce, T.}, \au{Ide, S.},
  \au{Becoulet, A.}, \au{Litaudon, X.}, \au{Isayama, A.}, \au{Hobirk, J.},
  \au{Wade, M.}, \au{Oikawa, T.}, \au{Prater, R.}, \au{Zvonkov, A.}, \au{Lloyd,
  B.}, \au{Suzuki, T.}, \au{Barbato, E.}, \au{Bonoli, P.}, \au{Phillips, C.},
  \au{Vdovin, V.}, \au{Joffrin, E.}, \au{Casper, T.}, \au{Ferron, J.},
  \au{Mazon, D.}, \au{Moreau, D.}, \au{Bundy, R.}, \au{Kessel, C.},
  \au{Fukuyama, A.}, \au{Hayashi, N.}, \au{Imbeaux, F.}, \au{Murakami, M.},
  \au{Polevoi, A.} \& \au{John, H.~S.}} \yr{2007}  \at{Chapter 6: Steady state
  operation}.  \jt{Nuclear Fusion}  \bvol{47},  \pg{S285--S336}.

\bibitem[Guo {\em et~al.\/}(2009)Guo, Chen \& Zonca]{ZGuoPRL2009}
{\sc \au{Guo, Z.}, \au{Chen, L.} \& \au{Zonca, F.}} \yr{2009}  \at{Radial
  spreading of drift-wave \& zonal-flow turbulence via soliton formation}.
  \jt{Phys. Rev. Lett.}  \bvol{103},  \pg{055002}.

\bibitem[Hahm {\em et~al.\/}(1999)Hahm, Beer, Lin, Hammett, Lee \&
  Tang]{TSHahmPoP1999}
{\sc \au{Hahm, T.~S.}, \au{Beer, M.~A.}, \au{Lin, Z.}, \au{Hammett, G.~W.},
  \au{Lee, W.~W.} \& \au{Tang, W.~M.}} \yr{1999}  \at{Shearing rate of
  time-dependent e x b flow}.  \jt{Physics of Plasmas}  \bvol{6}~(3),
  \pg{922}.

\bibitem[Hahm \& Chen(1995)]{TSHahmPRL1995}
{\sc \au{Hahm, T.~S.} \& \au{Chen, L.}} \yr{1995}  \at{Nonlinear saturation of
  toroidal alfv\'en eigenmodes via ion compton scattering}.  \jt{Phys. Rev.
  Lett.}  \bvol{74},  \pg{266}.

\bibitem[Huang {\em et~al.\/}(2020)Huang, Garofalo, Qian, Gong, Ding, Varela,
  Chen, Guo, Li, Wu, Pan, Ren, Zhang, Lao, Holcomb, McClenaghan, Weisberg,
  Chan, Hyatt, Hu, Li, Ferron, McKee, Pinsker, Rhodes, Staebler, Spong \&
  Yan]{JHuangNF2020}
{\sc \au{Huang, J.}, \au{Garofalo, A.}, \au{Qian, J.}, \au{Gong, X.}, \au{Ding,
  S.}, \au{Varela, J.}, \au{Chen, J.}, \au{Guo, W.}, \au{Li, K.}, \au{Wu, M.},
  \au{Pan, C.}, \au{Ren, Q.}, \au{Zhang, B.}, \au{Lao, L.}, \au{Holcomb, C.},
  \au{McClenaghan, J.}, \au{Weisberg, D.}, \au{Chan, V.}, \au{Hyatt, A.},
  \au{Hu, W.}, \au{Li, G.}, \au{Ferron, J.}, \au{McKee, G.}, \au{Pinsker, R.},
  \au{Rhodes, T.}, \au{Staebler, G.}, \au{Spong, D.} \& \au{Yan, Z.}} \yr{2020}
   \at{Progress in extending high poloidal beta scenarios on diii-d towards a
  steady-state fusion reactor and impact of energetic particles}.  \jt{Nuclear
  Fusion}  \bvol{60}~(12),  \pg{126007}.

\bibitem[Lauber(2013)]{PLauberPR2013}
{\sc \au{Lauber, P.}} \yr{2013}  \at{Super-thermal particles in hot
  plasmas-kinetic models, numerical solution strategies, and comparison to
  tokamak experiments}.  \jt{Physics Reports}  \bvol{533}~(2),  \pg{33 -- 68}.

\bibitem[Lin {\em et~al.\/}(1998)Lin, Hahm, Lee, Tang \&
  White]{ZLinScience1998}
{\sc \au{Lin, Z.}, \au{Hahm, T.~S.}, \au{Lee, W.~W.}, \au{Tang, W.~M.} \&
  \au{White, R.~B.}} \yr{1998}  \at{Turbulent transport reduction by zonal
  flows: Massively parallel simulations}.  \jt{Science}  \bvol{281}~(5384),
  \pg{1835--1837}.

\bibitem[Qiu {\em et~al.\/}(2009)Qiu, Chen \& Zonca]{ZQiuPPCF2009}
{\sc \au{Qiu, Z.}, \au{Chen, L.} \& \au{Zonca, F.}} \yr{2009}
  \at{Collisionless damping of short wavelength geodesic acoustic modes}.
  \jt{Plasma Physics and Controlled Fusion}  \bvol{51}~(1),  \pg{012001}.

\bibitem[Qiu {\em et~al.\/}(2016{\natexlab{{\em a\/}}})Qiu, Chen \&
  Zonca]{ZQiuPoP2016}
{\sc \au{Qiu, Z.}, \au{Chen, L.} \& \au{Zonca, F.}} \yr{2016{\natexlab{{\em
  a\/}}}}  \at{Effects of energetic particles on zonal flow generation by
  toroidal alfv\'en eigenmode}.  \jt{Physics of Plasmas}  \bvol{23}~(9),
  \pg{090702}.

\bibitem[Qiu {\em et~al.\/}(2016{\natexlab{{\em b\/}}})Qiu, Chen \&
  Zonca]{ZQiuNF2016}
{\sc \au{Qiu, Z.}, \au{Chen, L.} \& \au{Zonca, F.}} \yr{2016{\natexlab{{\em
  b\/}}}}  \at{Fine radial structure zonal flow excitation by beta-induced
  alfv\'en eigenmode}.  \jt{Nuclear Fusion}  \bvol{56}~(10),  \pg{106013}.

\bibitem[Qiu {\em et~al.\/}(2017)Qiu, Chen \& Zonca]{ZQiuNF2017}
{\sc \au{Qiu, Z.}, \au{Chen, L.} \& \au{Zonca, F.}} \yr{2017}  \at{Nonlinear
  excitation of finite-radial-scale zonal structures by toroidal alfv{\'e}n
  eigenmode}.  \jt{Nuclear Fusion}  \bvol{57}~(05),  \pg{056017}.

\bibitem[Qiu {\em et~al.\/}(2019{\natexlab{{\em a\/}}})Qiu, Chen \&
  Zonca]{ZQiuNF2019a}
{\sc \au{Qiu, Z.}, \au{Chen, L.} \& \au{Zonca, F.}} \yr{2019{\natexlab{{\em
  a\/}}}}  \at{Gyrokinetic theory of the nonlinear saturation of a toroidal
  alfv{\'{e}}n eigenmode}.  \jt{Nuclear Fusion}  \bvol{59}~(6),  \pg{066024}.

\bibitem[Qiu {\em et~al.\/}(2018{\natexlab{{\em a\/}}})Qiu, Chen, Zonca \&
  Chen]{ZQiuPRL2018}
{\sc \au{Qiu, Z.}, \au{Chen, L.}, \au{Zonca, F.} \& \au{Chen, W.}}
  \yr{2018{\natexlab{{\em a\/}}}}  \at{Nonlinear decay and plasma heating by a
  toroidal alfv\'en eigenmode}.  \jt{Phys. Rev. Lett.}  \bvol{120}~(13),
  \pg{135001}.

\bibitem[Qiu {\em et~al.\/}(2018{\natexlab{{\em b\/}}})Qiu, Chen, Zonca \&
  Chen]{ZQiuIAEAFEC2018}
{\sc \au{Qiu, Z.}, \au{Chen, L.}, \au{Zonca, F.} \& \au{Chen, W.}}
  \yr{2018{\natexlab{{\em b\/}}}} Nonlinear decay and plasma heating by
  toroidal alfv\'en eigenmodes.  \bt{In {\em 27th IAEA FEC\/}},  \pg{pp.
  TH/P2--6}. Ahmedabad, India.

\bibitem[Qiu {\em et~al.\/}(2019{\natexlab{{\em b\/}}})Qiu, Chen, Zonca \&
  Chen]{ZQiuNF2019b}
{\sc \au{Qiu, Z.}, \au{Chen, L.}, \au{Zonca, F.} \& \au{Chen, W.}}
  \yr{2019{\natexlab{{\em b\/}}}}  \at{Nonlinear excitation of a geodesic
  acoustic mode by toroidal alfv{\'{e}}n eigenmodes and the impact on plasma
  performance}.  \jt{Nuclear Fusion}  \bvol{59}~(6),  \pg{066031}.

\bibitem[Rosenbluth \& Hinton(1998)]{MRosenbluthPRL1998}
{\sc \au{Rosenbluth, M.~N.} \& \au{Hinton, F.~L.}} \yr{1998}  \at{Poloidal flow
  driven by ion-temperature-gradient turbulence in tokamaks}.  \jt{Phys. Rev.
  Lett.}  \bvol{80},  \pg{724--727}.

\bibitem[Sharapov {\em et~al.\/}(2002)Sharapov, Alper, Berk, Borba, Breizman,
  Challis, Fasoli, Hawkes, Hender, Mailloux, Pinches, Testa \& work
  programme]{SSharapovPoP2002}
{\sc \au{Sharapov, S.~E.}, \au{Alper, B.}, \au{Berk, H.~L.}, \au{Borba, D.~N.},
  \au{Breizman, B.~N.}, \au{Challis, C.~D.}, \au{Fasoli, A.}, \au{Hawkes,
  N.~C.}, \au{Hender, T.~C.}, \au{Mailloux, J.}, \au{Pinches, S.~D.},
  \au{Testa, D.} \& \au{work programme, E.-J.}} \yr{2002}  \at{Alfv\'en wave
  cascades in a tokamak}.  \jt{Physics of Plasmas}  \bvol{9},  \pg{2027}.

\bibitem[Sharapov {\em et~al.\/}(2001)Sharapov, Testa, Alper, Borba, Fasoli,
  Hawkes, Heeter, Mantsinen \& Hellermann]{SSharapovPLA2001}
{\sc \au{Sharapov, S.~E.}, \au{Testa, D.}, \au{Alper, B.}, \au{Borba, D.~N.},
  \au{Fasoli, A.}, \au{Hawkes, N.~C.}, \au{Heeter, R.~F.}, \au{Mantsinen,
  M.~J.} \& \au{Hellermann, M. G.~V.}} \yr{2001}  \at{Mhd spectroscopy through
  detecting toroidal alfv\'en eigenmodes and alfv\'en wave cascades}.
  \jt{Physics Letters A}  \bvol{289},  \pg{127}.

\bibitem[Shi {\em et~al.\/}(2020)Shi, Qiu, Chen, Wang, Shi, Yu, Yang, Zhong,
  Jiang, Ji, Yang, Xu \& Duan]{PShiNF2020}
{\sc \au{Shi, P.}, \au{Qiu, Z.}, \au{Chen, W.}, \au{Wang, Z.}, \au{Shi, Z.},
  \au{Yu, L.}, \au{Yang, Z.}, \au{Zhong, W.}, \au{Jiang, M.}, \au{Ji, X.},
  \au{Yang, Q.}, \au{Xu, M.} \& \au{Duan, X.}} \yr{2020}  \at{Thermal ions heat
  transport induced by reversed shear alfv\'en eigenmode on the hl-2a tokamak}.
   \jt{Nuclear Fusion}  \bvol{60},  \pg{064001}.

\bibitem[Todo {\em et~al.\/}(2010)Todo, Berk \& Breizman]{YTodoNF2010}
{\sc \au{Todo, Y.}, \au{Berk, H.} \& \au{Breizman, B.}} \yr{2010}
  \at{Nonlinear magnetohydrodynamic effects on alfv{\'e}n eigenmode evolution
  and zonal flow generation}.  \jt{Nuclear Fusion}  \bvol{50}~(8),
  \pg{084016}.

\bibitem[Tomabechi {\em et~al.\/}(1991)Tomabechi, Gilleland, Sokolov, Toschi \&
  the ITER~Team]{KTomabechiNF1991}
{\sc \au{Tomabechi, K.}, \au{Gilleland, J.}, \au{Sokolov, Y.}, \au{Toschi, R.}
  \& \au{the ITER~Team}} \yr{1991}  \at{Iter conceptual design}.  \jt{Nuclear
  Fusion}  \bvol{31}~(6),  \pg{1135}.

\bibitem[Wang {\em et~al.\/}(2018)Wang, Qiu, Zonca, Briguglio, Fogaccia, Vlad
  \& Wang]{TWangPoP2018}
{\sc \au{Wang, T.}, \au{Qiu, Z.}, \au{Zonca, F.}, \au{Briguglio, S.},
  \au{Fogaccia, G.}, \au{Vlad, G.} \& \au{Wang, X.}} \yr{2018}  \at{Shear
  alfv\'en fluctuation spectrum in divertor tokamak test facility plasmas}.
  \jt{Physics of Plasmas}  \bvol{25}~(6),  \pg{062509}.

\bibitem[Wang {\em et~al.\/}(2020)Wang, Qiu, Zonca, Briguglio \&
  Vlad]{TWangNF2020}
{\sc \au{Wang, T.}, \au{Qiu, Z.}, \au{Zonca, F.}, \au{Briguglio, S.} \&
  \au{Vlad, G.}} \yr{2020}  \at{Dynamics of reversed shear alfv{\'{e}}n
  eigenmode and energetic particles during current ramp-up}.  \jt{Nuclear
  Fusion}  \bvol{60}~(12),  \pg{126032}.

\bibitem[Wang {\em et~al.\/}(2019)Wang, Wang, Briguglio, Qiu, Vlad \&
  Zonca]{TWangPoP2019}
{\sc \au{Wang, T.}, \au{Wang, X.}, \au{Briguglio, S.}, \au{Qiu, Z.}, \au{Vlad,
  G.} \& \au{Zonca, F.}} \yr{2019}  \at{Nonlinear dynamics of shear alfv\'en
  fluctuations in divertor tokamak test facility plasmas}.  \jt{Physics of
  Plasmas}  \bvol{26}~(1),  \pg{012504}.

\bibitem[Wei {\em et~al.\/}(2021)Wei, Wang, Shi, Chen, Chen \&
  Qiu]{SWeiCPL2021}
{\sc \au{Wei, S.}, \au{Wang, Y.}, \au{Shi, P.}, \au{Chen, W.}, \au{Chen, N.} \&
  \au{Qiu, Z.}} \yr{2021}  \at{Nonlinear coupling of reversed shear alfv\'en
  eigenmode and toroidal alfv\'en eigenmode during current ramp}.  \jt{Chinese
  Physics Letters}  \bvol{38},  \pg{035201}.

\bibitem[White {\em et~al.\/}(2010{\natexlab{{\em a\/}}})White, Gorelenkov,
  Heidbrink \& Zeeland]{RWhitePoP2010}
{\sc \au{White, R.~B.}, \au{Gorelenkov, N.}, \au{Heidbrink, W.~W.} \&
  \au{Zeeland, M. A.~V.}} \yr{2010{\natexlab{{\em a\/}}}}  \at{Beam
  distribution modification by alfv\'en modes}.  \jt{Physics of Plasmas}
  \bvol{17},  \pg{056107}.

\bibitem[White {\em et~al.\/}(2010{\natexlab{{\em b\/}}})White, Gorelenkov,
  Heidbrink \& Zeeland]{RWhitePPCF2010}
{\sc \au{White, R.~B.}, \au{Gorelenkov, N.}, \au{Heidbrink, W.~W.} \&
  \au{Zeeland, M. A.~V.}} \yr{2010{\natexlab{{\em b\/}}}}  \at{Particle
  distribution modification by low amplitude modes}.  \jt{Plasma Physics and
  Controlled Fusion}  \bvol{52},  \pg{045012}.

\bibitem[Zonca {\em et~al.\/}(2002)Zonca, Briguglio, Chen, Dettrick, Fogaccia,
  Testa \& Vlad]{FZoncaPoP2002}
{\sc \au{Zonca, F.}, \au{Briguglio, S.}, \au{Chen, L.}, \au{Dettrick, S.},
  \au{Fogaccia, G.}, \au{Testa, D.} \& \au{Vlad, G.}} \yr{2002}  \at{Energetic
  particle mode stability in tokamaks with hollow q-profiles}.  \jt{Physics of
  Plasmas}  \bvol{9}~(12),  \pg{4939--4956}.

\bibitem[Zonca \& Chen(2008)]{FZoncaEPL2008}
{\sc \au{Zonca, F.} \& \au{Chen, L.}} \yr{2008}  \at{Radial structures and
  nonlinear excitation of geodesic acoustic modes}.  \jt{Europhys. Lett.}
  \bvol{83}~(3),  \pg{35001}.

\bibitem[Zonca {\em et~al.\/}(2015{\natexlab{{\em a\/}}})Zonca, Chen,
  Briguglio, Fogaccia, Milovanov, Qiu, Vlad \& Wang]{FZoncaPPCF2015}
{\sc \au{Zonca, F.}, \au{Chen, L.}, \au{Briguglio, S.}, \au{Fogaccia, G.},
  \au{Milovanov, A.~V.}, \au{Qiu, Z.}, \au{Vlad, G.} \& \au{Wang, X.}}
  \yr{2015{\natexlab{{\em a\/}}}}  \at{Energetic particles and multi-scale
  dynamics in fusion plasmas}.  \jt{Plasma Physics and Controlled Fusion}
  \bvol{57}~(1),  \pg{014024}.

\bibitem[Zonca {\em et~al.\/}(2015{\natexlab{{\em b\/}}})Zonca, Chen,
  Briguglio, Fogaccia, Vlad \& Wang]{FZoncaNJP2015}
{\sc \au{Zonca, F.}, \au{Chen, L.}, \au{Briguglio, S.}, \au{Fogaccia, G.},
  \au{Vlad, G.} \& \au{Wang, X.}} \yr{2015{\natexlab{{\em b\/}}}}
  \at{Nonlinear dynamics of phase space zonal structures and energetic particle
  physics in fusion plasmas}.  \jt{New Journal of Physics}  \bvol{17}~(1),
  \pg{013052}.

\bibitem[Zonca {\em et~al.\/}(2021)Zonca, Chen, Falessi \& Qiu]{FZoncaJPCS2021}
{\sc \au{Zonca, F.}, \au{Chen, L.}, \au{Falessi, M.~V.} \& \au{Qiu, Z.}} \at{
  \yr{2021} } \jt{Journal of Physics: Conference Series}  \bvol{1785},
  \pg{012005}.

\bibitem[Zonca {\em et~al.\/}(1995)Zonca, Romanelli, Vlad \&
  Kar]{FZoncaPRL1995}
{\sc \au{Zonca, F.}, \au{Romanelli, F.}, \au{Vlad, G.} \& \au{Kar, C.}}
  \yr{1995}  \at{Nonlinear saturation of toroidal alfv{\'e}n eigenmodes}.
  \jt{Phys. Rev. Lett.}  \bvol{74}~(5),  \pg{698}.

\end{thebibliography}


\begin{thebibliography}{14}
\expandafter\ifx\csname natexlab\endcsname\relax\def\natexlab#1{#1}\fi

\bibitem[Batchelor(1971)]{Batchelor59}
{\sc Batchelor, G.~K.} 1971 Small-scale variation of convected quantities like
  temperature in turbulent fluid. part 1. general discussion and the case of
  small conductivity. {\em J.~Fluid Mech.\/} {\bf 5}, 113--133.

\bibitem[Brownell \& Su(2004)]{Brownell04}
{\sc Brownell, C.~J. \& Su, L.~K.} 2004 Planar measurements of differential
  diffusion in turbulent jets. {\em AIAA Paper 2004-2335\/}.

\bibitem[Brownell \& Su(2007)]{Brownell07}
{\sc Brownell, C.~J. \& Su, L.~K.} 2007 Scale relations and spatial spectra in
  a differentially diffusing jet. {\em AIAA Paper 2007-1314\/}.

\bibitem[Dennis(1985)]{Dennis85}
{\sc Dennis, S. C.~R.} 1985 {Compact explicit finite difference approximations
  to the Navier--Stokes equation}. In {\em Ninth Intl Conf. on Numerical
  Methods in Fluid Dynamics\/} (ed. Soubbaramayer \& J.~P. Boujot), {\em
  Lecture Notes in Physics\/}, vol. 218, pp. 23--51. Springer.

\bibitem[Hwang \& Tuck(1970)]{Hwang70}
{\sc Hwang, L.-S. \& Tuck, E.~O.} 1970 On the oscillations of harbours of
  arbitrary shape. {\em J.~Fluid Mech.\/} {\bf 42}, 447--464.

\bibitem[Koch(1983)]{Koch83}
{\sc Koch, W.} 1983 Resonant acoustic frequencies of flat plate cascades. {\em
  J.~Sound Vib.\/} {\bf 88}, 233--242.

\bibitem[Lee(1971)]{Lee71}
{\sc Lee, J.-J.} 1971 Wave-induced oscillations in harbours of arbitrary
  geometry. {\em J.~Fluid Mech.\/} {\bf 45}, 375--394.

\bibitem[Linton \& Evans(1992)]{Linton92}
{\sc Linton, C.~M. \& Evans, D.~V.} 1992 The radiation and scattering of
  surface waves by a vertical circular cylinder in a channel. {\em Phil.\
  Trans.\ R. Soc.\ Lond.\/} {\bf 338}, 325--357.

\bibitem[Martin(1980)]{Martin80}
{\sc Martin, P.~A.} 1980 On the null-field equations for the exterior problems
  of acoustics. {\em Q.~J. Mech.\ Appl.\ Maths\/} {\bf 33}, 385--396.

\bibitem[Miller(1991)]{Miller91}
{\sc Miller, P.~L.} 1991 Mixing in high schmidt number turbulent jets. PhD
  thesis, California Institute of Technology.

\bibitem[Rogallo(1981)]{Rogallo81}
{\sc Rogallo, R.~S.} 1981 Numerical experiments in homogeneous turbulence. {\em
  Tech. Rep.\/} 81835. NASA Tech.\ Mem.

\bibitem[Ursell(1950)]{Ursell50}
{\sc Ursell, F.} 1950 Surface waves on deep water in the presence of a
  submerged cylinder i. {\em Proc.\ Camb.\ Phil.\ Soc.\/} {\bf 46}, 141--152.

\bibitem[{van Wijngaarden}(1968)]{Wijngaarden68}
{\sc {van Wijngaarden}, L.} 1968 On the oscillations near and at resonance in
  open pipes. {\em J.~Engng Maths\/} {\bf 2}, 225--240.

\bibitem[Worster(1992)]{Worster92}
{\sc Worster, M.~G.} 1992 {The dynamics of mushy layers}. In {\em In
  Interactive dynamics of convection and solidification\/} (ed. S.~H. Davis,
  H.~E. Huppert, W.~Muller \& M.~G. Worster), pp. 113--138. Kluwer.

\end{thebibliography}

\end{document}